# A New Technique to Backup and Restore DBMS using XML and .NET Technologies

Seifedine Kadry [1,2,3], Mohamad Smaili [2], Hussam Kassem [1,2], Hassan Hayek [3]

[1] *Lebanese University, ISAE-CNAM, Lebanon*
[2] *Lebanese University, Faculty of Science, Lebanon*
[3] *Arts, Sciences & Technology University in Lebanon*

skadry@gmail.com
*Address: LIU, P.O. Box 5, Jeb Janeen, Khyara, Bekaa, Lebanon*

*Abstract*— **In this paper, we proposed a new technique for backing up and restoring different Database Management Systems (DBMS). The technique is enabling to backup and restore a part of or the whole database using a unified interface using ASP.NET and XML technologies. It presents a Web Solution allowing the administrators to do their jobs from everywhere, locally or remotely. To show the importance of our solution, we have taken two case studies, oracle 11g and SQL Server 2008.**

## I. INTRODUCTION

All DBMS allow a full backup which take the whole database, saved it as a copy, and store in a specific location on the hard disk or on some storage devices (CD, DVD, Tapes…). The main questions that should be asked are: 1) How can we save some particular items of the database? 2) How can we backup only one table of a big database? Given that the size of the database is 300 G.B., do we backup the whole DBMS in order to restore later one table? How much time/space this consume?

On the other hand, most companies, and due to the different source of applications, they use more than one DBMS, like SQL Server [1], Oracle [2], Sybase, DB2 and others. Each one of these has a specific interface and tool for back up and restore. Does the database administrator have to learn and master all these tools? of course "No". To solve the problem, we propose a new technique uses only a unified and friendly user interface, which allows a solution backup and restore the database by choosing specific articles without learning how does each DBMS work, and also without opening them. The administrator can choose a specific article for Backup/Restore instead of the Backup/Restore the whole DBMS, and to avoid them to master any DBMS tool for Backup/Restore.

In this paper we achieve the following points:
- Partially backup and restore data articles from different DBMS.
- Provide a unique friendly user interface which allows a full backup and restore of different DBMS without any need to run these systems.
- The administrator can now backup and restore databases locally or remotely.

## II. RELATED WORK

All the known DBMS allow to backup and restore the available databases. Suppose we are working on a huge database, an error is occurred and corrupted; we failed to access this database. Because of this, **will we lose the whole data?** Each DBMS provides a backup and restores actions (jobs) which allow us to retrieve our old data at any time we need (**will we go back to zero?**).

Currently, the Backup/Restore DBMS too doesn't provide the backup/restore of a specific article of the DBMS [3, 4, 5]. Our proposed technique allows the user to partially backup and restore the following DBMS articles:
- Stored Procedure
- Functions
- Triggers
- Views
- Tables
- Records within tables.

After choosing the articles, the backup file will be saved as an XML document for its simplicity and reliability.

## III. BACKUP TYPES

In information technology, backup refers to making copies of data so that these additional copies may be used to restore the original after a data loss occur [6]. These additional copies are typically called "backups". Backups are useful primarily for two purposes. The first is to restore a state following a disaster (called disaster recovery), and the second is to restore small numbers of files after they have been accidentally deleted or corrupted.

You can perform different types of backups. For instance, in SQL Server 2008, three types of backup are available: Full, Differential, and Transaction log Backups [7]. In Oracle 11g Database, 2 types of backup are available: Cold (offline), and Hot (Online) backups [8].

### A. SQL Server Types

*Full Backup*

Full database backups are the default and the starting point for all other types of backups. A full database backup captures the entire database, including all entries in the transaction log-excluding any unallocated extents in the files. Pages are read directly from disk to increase the speed of the operation.

*Differential Backup*

Differential backups capture all the data that has changed since the last full database backup. Differential backups will increase the speed of the backup operation as well as the restore. Because only the changed or newly allocated extents are captured, differential backups are faster and smaller than full database backups.

*Transaction Log Backup*

Transaction log backups serially capture modifications to the database. Backups of the transaction log provide a history of the transactions that have taken place within the database. The backups of the log are then used in the recovery process to restore the database fully, to a point in time. Transaction log backups are applied to recover a database by rolling forward (redoing) any committed changes not reflected in the database and rolling back (undoing) uncommitted transactions. Log backups are smaller and are taken more frequently than full or differential backups.

*B. Oracle Backup Types*

*Cold Backup*

A cold backup is when the database is not running - i.e. users are not logged on - hence no activity going on and easier to backup. This is also known as an offline backup, which can only give a read-consistent copy but doesn't handle active transactions. You must ensure that all redo logs archived during the backup process are also backed up. The hot backup differs from the cold backup in that only sections of the database are backed up at one time

*Hot Backup*

A hot backup would have to be taken if your database is mission critical, i.e. it has to run 24 hours a day, 7 days a week (Banks and ATM machines, communications…). In this case you will have to perform an online or a hot backup.

IV. TECHNOLOGIES AND TECHNIQUES USED IN OUR SOLUTION

*XML*

Some people consider eXtensible Markup Language (XML) to be one of the most important new technologies appeared in the parade of technologies developed to support the World Wide Web. It provides a standard way to represent information, which allows that information to be stored and interchanged among any Internet-connected devices. It also allows any number of different software systems to manipulate that information [9, 10].

We can observe that XML is easy to write because of the ways in which the data is stored and structured, and because of the simplicity of interaction between .Net technology and XML, we use XML technology to hold our backup files; i.e. the backed up database articles are saved within XML document.

*.NET TECHNOLGY*

ASP.NET was build by Microsoft from one of their biggest technologies. Web programmers can make use of any encoding language they want to write ASP.NET, from Perl to C Sharp (C#) and of course VB.NET, and a few extra languages unspoken with the .NET technology.

ASP.NET is a server side scripting technology that enables scripts (embedded in web pages) to be executed by an internet or web server [11, 12].

We have used .NET Technology in our proposed technique, because, one of the most important ASP.Net classes is the "Dataset" class which can hold your database tables and can read/write from/into XML documents [13].

Dataset can be defined as:
```
Dim ds as new Dataset()
```

Tables can be added into the dataset as following:
```
ds.tables.add(my_index)
```

Dataset can hold big number of these data tables, these tables are exported into XML document using the following piece of code:
```
Dim xmlDoc As New FileStream("my path",
FileMode.CreateNew)
ds.WriteXml(xmlDoc)
```

This will create a new XML document that contains data tables of the data set, and then save it at "my path".

To import data from XML document and place them in the data set, you have to use the method ReadXml
```
Dim xmlDoc As New FileStream(("my path",
FileMode.Open)
DS.ReadXml(xmlDoc)
```

*Microsoft SQL Server 2008*

It is used as a main DBMS in our proposed technique:
- Users: contains users' names and passwords allow users to login and access our site (Table 1).

Table 1: Users Table

| ID | UserName | Password |
|----|----------|----------|
| 19 | user1 | 123456 |
| 20 | user20 | pswrd20 |
| 21 | user21 | pswrd21 |

- Statements: all queries used in our software are saved in this table. The idea here is to parameterize the query to be able to add later any new DBMS easily. The current software supports two DBMS: SQL server 2005 and Oracle 10g. Suppose we were asked later to make the system support new DMBS as Sybase, all what we have to do is: making few changes in the code, and adding the needed query into the statements table, this will solve the problem (Table 2).

**Note**: To connect from .NET to any DBMS, we can use some predefined classes, like the System.Data. SqlClient class for SQL server, then, we define a connection string (ConnString) as string, and the connection Conn as new SQLConnection(ConnString).

Using Conn.open() is enough to start accessing the MSSQL server DBMS.

Similarly, we can access Oracle DBMS: Replacing System.Data. SqlClient by System.Data.OracleClient, and Conn as new SQLConnection(ConnString) by Conn as new OracleConnection(ConnString), then Conn.open()

## V. THE PROPOSED TECHNIQUE

A new design and implementation for backing up different DBMS is developed. The idea is based on the creation of a unified interface using ASP.NET and XML technology.

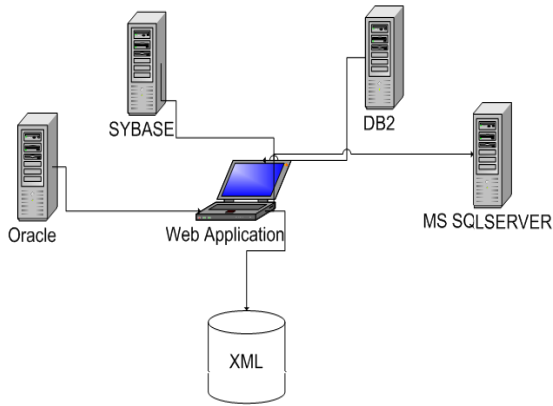

Fig. 1: Solution Diagram

The technique is applied to two well known DBMS: MSSQL Server 2008 and Oracle 11g. Taking into consideration the need of backups at some urgent times (e.g. fire in the building), where the database administrators cannot reach the server. How do we backup the database to a secured place? Simply the solution should be using the internet. Therefore, the solution is built as a website. The database administrators can backup and restore database regardless of his place. The proposed solution may be summarized by the following Fig. 1, which is 3-tier application architecture.

### Description of the Application
#### A. Startup page

The user is asked to enter a user name and password, if they are both valid, he will take permission to access other pages. In the case they are not valid, a message box pops up and ask the user to enter a correct user name and password.

The website supports two DBMS (others can be easily added later): SQL server 2008 and Oracle 11g. Users are free to choose one of these two DBMSs in order to do the backup or restore operation.

The "Log out" button is present on all pages except for "Log in" page. Once a user click it, this button call a method "Disconnect All" which kills all the opened connections with all available databases, and then forward us to the "Log in" page.

#### B. Connection Page

This page is divided into two Frames:

*Connection Frame:* is used to add test the connection components. Three fields and a button are within this frame: Server Name, User Name, Password, and Test button. Server Name is retrieved dynamically by the system. The user should select one of these servers, and insert the correct user name and password (for connection). The button Test is used to test if the connection is successfully built or not (Fig. 2 and Fig. 3).

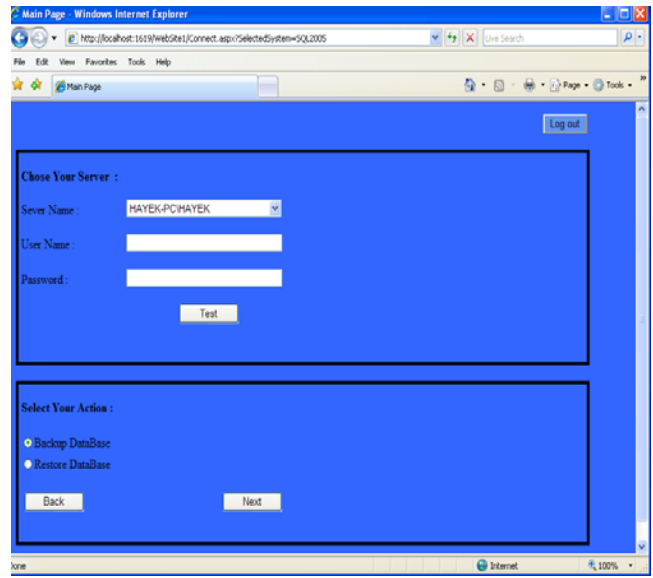

Fig.2: Connection Page

Table 2: Statements Tables

| ID | DB_ID | DB_Type | Stmnt_Specification | Query |
|----|-------|---------|---------------------|-------|
| 1 | 0 | SQL2005 | Get_All _DataBases | SELECT * FROM sysdatabases where dbid>4 |
| 2 | 0 | SQL2005 | Get_All _Tables | SELECT * FROM sys.tables where name <>'sysdiagrams' |
| 3 | 0 | SQL2005 | Get_All _StoredProced... | SELECT * FROM sysobjects , syscomments where sysobjects.id = sysc... |
| 4 | 0 | SQL2005 | Get_All_Views | SELECT * FROM  sysobjects , syscomments where sysobjects.id = sys... |
| 5 | 0 | SQL2005 | Get_All_Functions | SELECT * FROM  sysobjects , syscomments where sysobjects.id = sys... |
| 6 | 0 | SQL2005 | Get_All_Triggers | SELECT * FROM  sysobjects , syscomments where sysobjects.id = sys... |
| 7 | 0 | SQL2005 | Get_Selected_StoredP... | SELECT * FROM  sysobjects , syscomments where sysobjects.id = sysc... |
| 8 | 0 | SQL2005 | Get_Selected_Views | SELECT * FROM  sysobjects , syscomments where sysobjects.id = sys... |
| 9 | 0 | SQL2005 | Get_Selected_DataBase | SELECT * FROM sysdatabases where dbid>4 And Name=' |
| 10 | 0 | SQL2005 | Get_Selected_Functions | SELECT * FROM  sysobjects , syscomments where sysobjects.id = sys... |
| 11 | 0 | SQL2005 | Get_Selected_Triggers | SELECT * FROM  sysobjects , syscomments where sysobjects.id = sys... |
| 12 | 0 | SQL2005 | Get_Selected_Tables | SELECT * FROM sys.tables where name <>'sysdiagrams' And Name = ' |
| 13 | 0 | SQL2005 | Get_All_Records | SELECT * FROM |
| 14 | 0 | SQL2005 | Delete_DataBase | DROP DATABASE |
| 15 | 0 | SQL2005 | Add_DataBase | CREATE  DATABASE |
| 16 | 0 | SQL2005 | Get_All_Attributes | SELECT sysobjects.name as table_name,syscolumns.name as column_n... |
| 17 | 0 | SQL2005 | Get_All_Keys | SELECT  T.TABLE_NAME,K.COLUMN_NAME FROM INFORMATION_SCH... |
| 18 | 0 | SQL2005 | Disconnect_All_Connec... | DECLARE SpidsToKill CURSOR FOR SELECT spid FROM master..sysproc... |
| 19 | 0 | SQL2005 | Get_DataBase_ID | Select dbid FROM sysdatabases where name =' |
| 20 | 1 | Oracle10g | Get_All_Servers | SELECT HOST_NAME FROM v$instance |
| 21 | 1 | Oracle10g | Get_All _DataBases | SELECT INSTANCE_NAME FROM v$instance |

**Note:** How to retrieve servers' name (instance) of SQL Server 2008?

Every time we install a new instance of SQL Server 2008, the name of this instance saved in the registry of the computer and this piece of code read them:

```
Imports Microsoft.SqlServer.Management.Smo
Dim mc As New Wmi.ManagedComputer()
DDLServers.Items.Clear()
For Each si As Wmi.ServerInstance In
mc.ServerInstances
DDLServers.Items.Add(si.Name)
Next
```

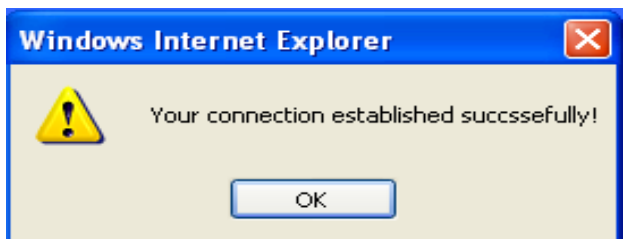

Fig. 3: Testing Successfully

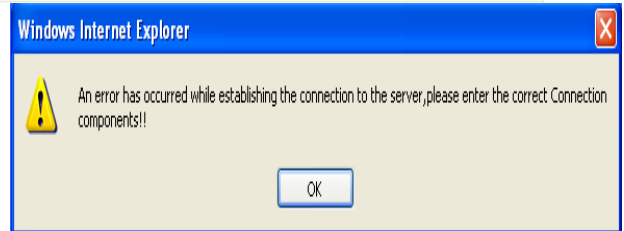

Fig. 4: Testing Failed

*Action Frame:* It is used to choose between Backup and restore. Two Categories of Backup are available:

- Partial Backup
- Full Backup

All the available databases in the selected server (in the previous page) will be retrieved and added in a drop down list. The user should choose a database in order to partially (or full) backup it. Once the database is selected, we must choose one of the two backup categories (Fig. 5).

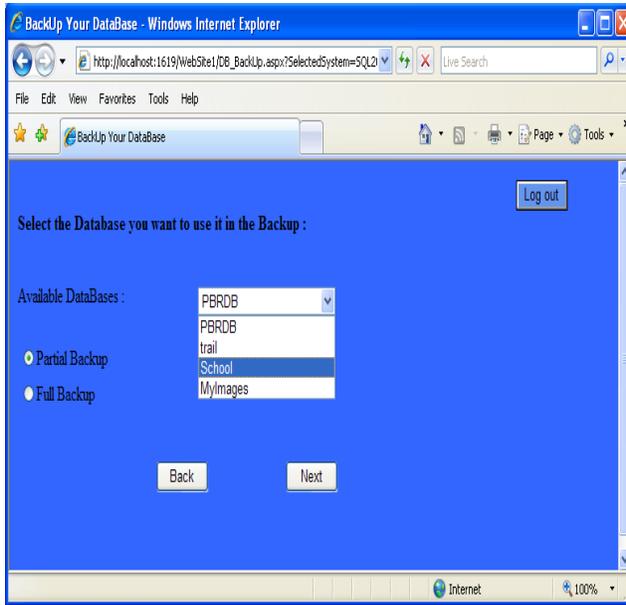

Fig. 5: Database Selection

## Partial Backup Page

Here, we test the part of the system that distinguishes it from other available systems. After choosing a database and selected "Partial Backup", a page opens (Fig. 6) in which we select the items to be backed up.

The connection string appears at the top of the page and below the label of the selected database appears.

Checking the combo boxes, we can select one of the stored procedure, Views, Functions, Triggers, Tables, and specific records to back them up.

From the figure above, we should note that:

- If no items exist (as in Functions and Triggers), the check box list becomes "gray".

- If a table is empty (has no entire records), the table appears as an empty table without any check boxes (as in the table "teacher").

- If a table is unchecked, then checking of any of its records is meaningless.

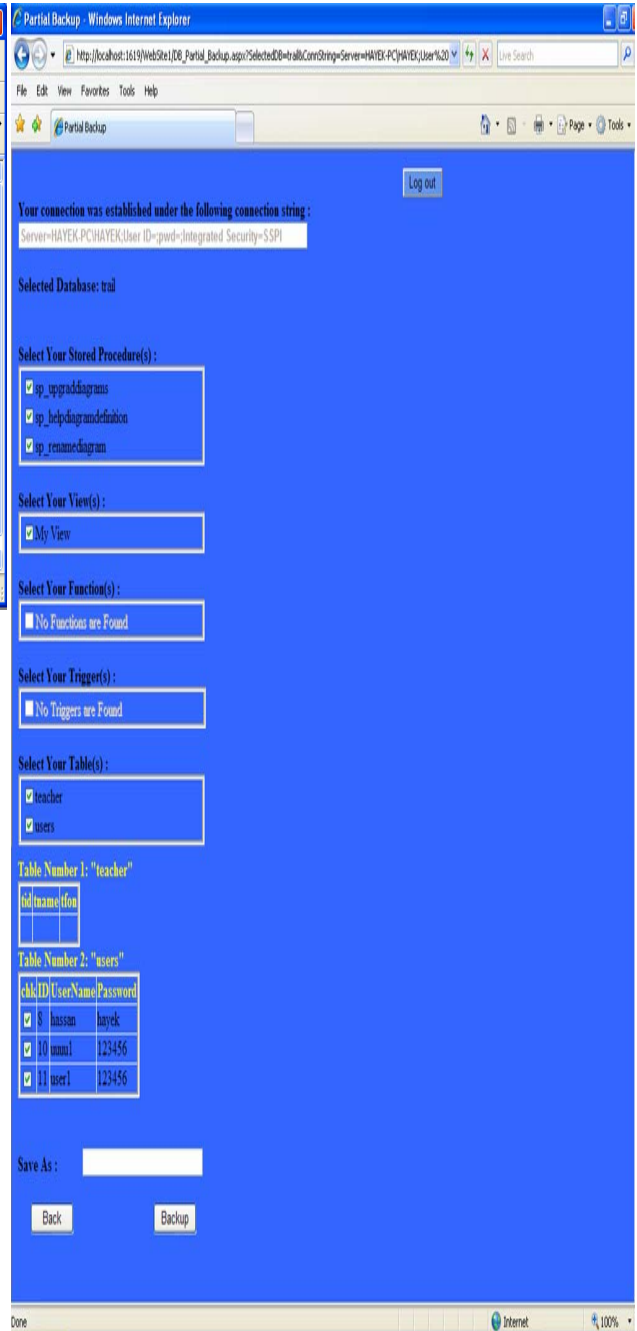

Fig. 6: Data Articles Selection

After selecting the items, the name of the backup file (XML document) must be inserted. This backup restored in the directory "C:\" (Fig. 7).

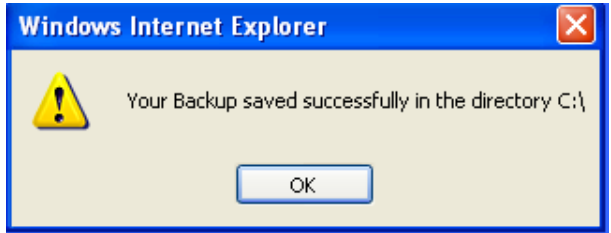

Fig. 7: Backup Completed

Note that if the XML file is damaged, another copy of the XML file saved in the directory C:\Backups\MyBackup.xml. An extra copy on some remote FTP server can be stored for maximum security.

The XML document contains:

- The selected DBMS: to prevent overlapping, if we backup a database from SQL server 2005, we have to restore it in SQL Server 2005 and not in Oracle 10g. Because of this, the header of the XML document should contain the DBMS :

```
<DataBase_Mangment_System>
<DBMS_name>SQL2008</DBMS_name>
</DataBase_Mangment_System>
```

- The selected database: and all its attributes (name, dbid, Sid, mode...).
- All the selected items: and there attributes (name, id, xtype, parent...).
- All the selected records (fields and values):

```
<Users>
<ID>1</ID>
<Username>user1</Username>
<Password>pswrd1</Password>
<ID>2</ID>
<Username>user2</Username>
<Password>pswrd2</Password>
</users>
```

Here, two users are selected from the table "users"(Fig. 8):

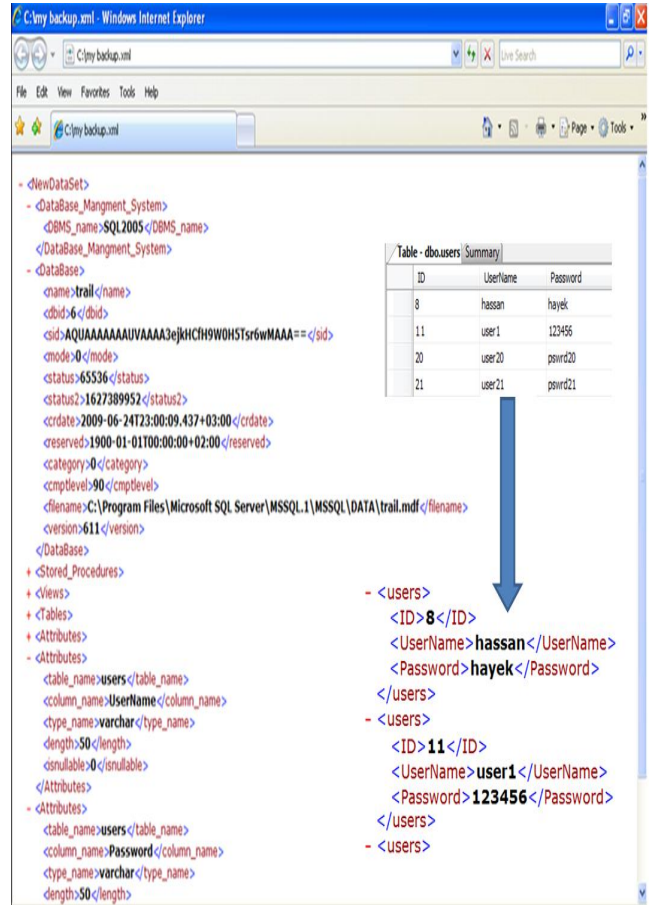

Fig. 8: Backup File - XML

Note that the backup support "concurrency control", i.e. if the user recently access the database, the backup will hold the last saved data without disconnecting the database.

*Full Backup Page*

The main goal of the full backup is similar to any Backup/Restore DBMS tool, which we can backup our databases from different DBMS. After choosing a database and selected "Full Backup", a page opens (Fig. 9) in which we backup our database. The backup file is also saved in the directory "C:\" under the name written in the textbox.

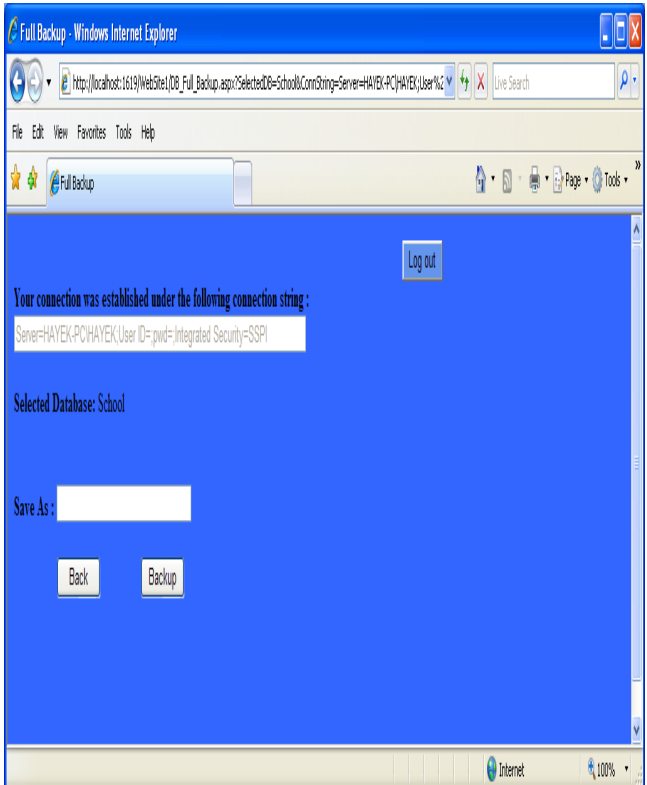

Fig. 9: Full Backup Page

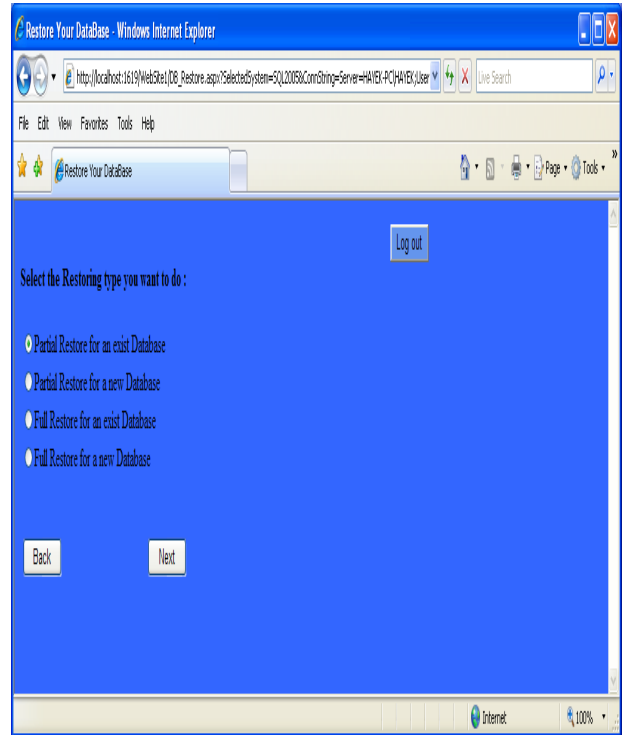

Fig. 10: Restore Categories

## Restore Pages

Four categories of restore are available:

- Partial Restore (exist) Page
- Partial Restore (new) Page
- Full Restore (exist) Page
- Full Restore (new) Page

As in backup, we can partially or fully restore the database. We can restore our database to an existing one, i.e. replace the existing database, or create a new database and restore it (Fig. 10).

## Partial Restore (exist) Page

All items within this database are automatically removed from it, and the database becomes null. The new items (Views, Triggers, Tables …) are transferred from the XML document into the empty database.

The header of the XML Document prevents overlapping between the DBMS, but, what will happen if the user chooses an XML document which is not a backup file?

Simply, the header of the XML document, and the tags of the database name guaranty the correctness of the backup file. If they do not exist, the XML is not a backup file. (Fig. 11, 12, and 13).

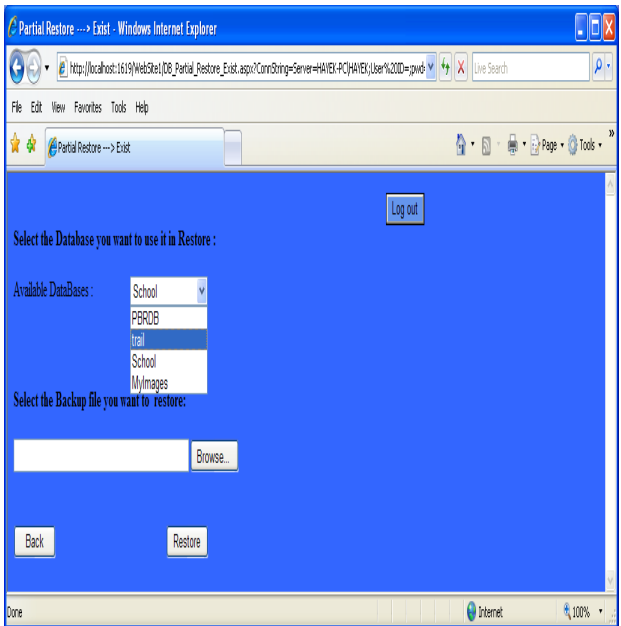

Fig. 11: Partially Restore into an exist Database

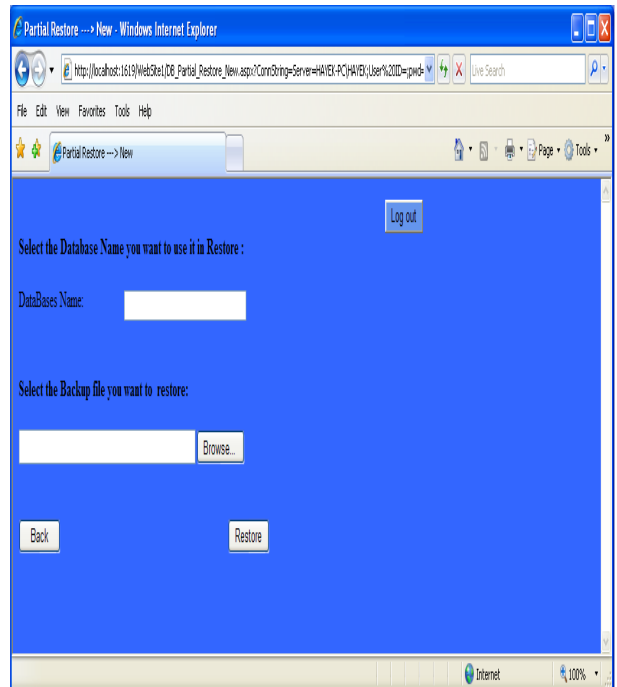

Fig. 14: Partially Restore into a new Database

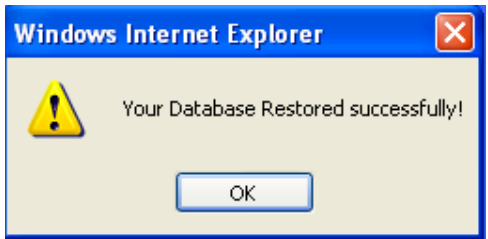

Fig. 12: Restore completed

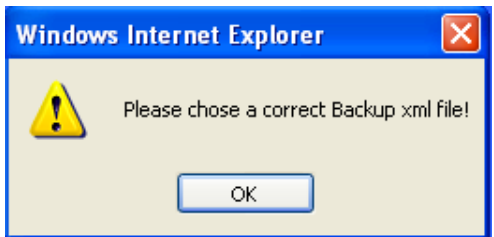

Fig. 13: Restore Failed

*Partial Restore (new) Page*

It is similar to the previous section, but, the only difference is that the user doesn't select any of the available databases to partially restore the XML document. It creates the new database that should hold the backup file (Fig. 14). The new database holds the name inserted by the user in the text box.

*Full Restore (exist) Page*

As mentioned before, the main goal of the full backup is the "unification" of the GUI, in which, we can backup our databases from different DBMSs.

Here, we are willing to achieve the same goal in the full restore. It's a simple interface that allows restoring databases in different DBMSs. We are restoring in an existing database (Fig. 11).

Also, it's a small query which can be executed inside the query analyzer. (See the following query in SQL Server 2005):

```
RESTORE DATABASE My_Database
FROM DISK = 'C:\MY_BACKUP.BAK'
```

*Full Restore (new) Page*

It restores the backup file into a new created database and holds the name entered by the user in the text box (Fig. 12).

Note that if the backup file already used by another database (restored before) then the restore procedure fails, and a message box informs the user (Fig. 15).

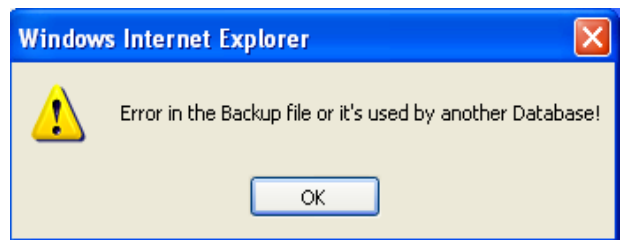

Fig. 15: Illegal Backup File

**Two problems:** We have faced two problems in our proposed technique: the full path and read/write images.

**Full path problem:** We tried to run it under the browsers Internet Explorer 8 (IE8) and Mozilla. It ran successfully, but the problem is in the Restore Page, when the user browses for a file to restore it.

In Internet Explorer 7 (IE7), when the user browse for a file from the "file upload" control (Fig. 16), the full path of this file can be accessed simply using the following code :
`myFileUpload.FileName`

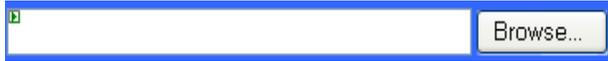

Fig. 16: Upload file Form

Suppose we want to fully restore a backup file from "c:\myfolder\My_Bakup.BAK", we simply write the following code:
```
RESTORE DATABASE My Database
FROM DISK = 'c:\myfolder\My_Bakup.BAK'
```

In IE7, myFileUpload.FileName returns the full path, i.e. c:\myfolder\My_Bakup.BAK, but in IE8 and Mozilla there was a new security "feature" that hides the real path of the selected file. It returned only the file name (My_Backup.BAK), so the restore statement became:
```
RESTORE DATABASE My Database
FROM DISK = 'My_Bakup.BAK', this caused the
```
failure of the restoring operation.

We solved the problem as follows:
When the user chooses a file to be restored, the server creates a new folder in the directory c:\ called "Temp_Restore", and then it takes a copy of the backup file and saves it in the created folder. Now the backup file is saved under the following path:
C:\Temp_Restore\My_Backup
Then we added hidden fields within the Restore pages. These hidden fields hold the full path, and then the problem is solved.
Note that, when system finishes the restoring operation, it automatically deletes the created copy.

*Images Problem*

In general, three methods are available to insert data into a database tables:

- Manually: by inserting the data from the table (Table 3).

Table 3: Users Table

| Table - dbo.users | Summary | |
|---|---|---|
| ID | UserName | Password |
| 1 | user1 | pswrd1 |
| 2 | user2 | pswrd2 |
| 3 | user3 | pswrd3 |
| ►* NULL | NULL | NULL |

- Queries: by executing some statements (insert, update …):
Insert into USERS values (123, 'user name 123', 'pswrd 123')
- From a GUI, using programming technologies (.Net, Net Beans…)

What will happen if the user wants to insert Images into the database? This can be done only from a GUI [14].
Images saved in the database as an Array of bytes, as we see in (Table 4).

Table 4: Images Table

| Table - dbo.Photographs | | | |
|---|---|---|---|
| ID | CategoryID | Name | Photograph |
| 1 | 3 | image1 | <Binary data> |
| 2 | 1 | image2 | <Binary data> |
| 3 | 1 | image3 | <Binary data> |
| 4 | 1 | image4 | <Binary data> |
| * NULL | NULL | NULL | NULL |

To explain this, we created a small VB.net windows application to clarify (Fig. 17).

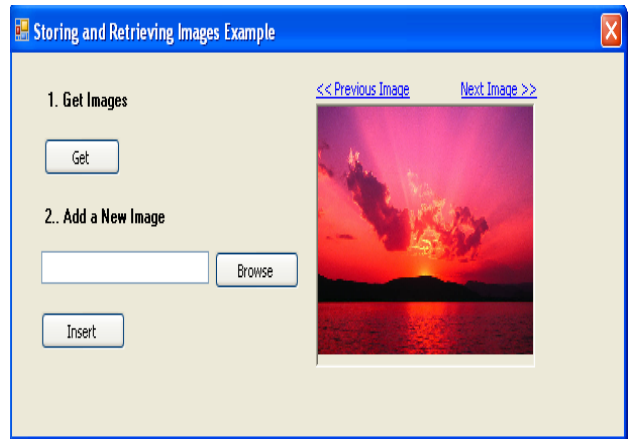

Fig. 17: Images Insertion

First, browse for a new image, then press insert. The system will change the image into a very big binary array (using a built in function in .Net "MemoryStream"), then the hexadecimal array is inserted into the table.

Now, let us make a selection statement and see what happens (Table 5).

Table 5: Images as Hexadecimal String

| | ID | CategoryID | Name | Photograph |
|---|---|---|---|---|
| 1 | 1 | 3 | image1 | 0xFFD8FFE0001044454640000120010063000000FFED01D1850686FA67F7B086F70020032C30003B42490003ED0A5267536F6C7574... |
| 2 | 2 | 1 | image2 | 0xFFD8FFE0001044454640000120010063000000FFED01D1850686FA67F7B086F70020032C30003B42490003ED0A5267536F6C7574... |
| 3 | 3 | 1 | image3 | 0xFFD8FFE0001044454640000120010063000000FFED01D1850686FA67F7B086F70020032C30003B42490003ED0A5267536F6C7574... |
| 4 | 4 | 1 | image4 | 0xFFD8FFE0001044454640000120010063000000FFED01D1850686FA67F7B086F70020032C30003B42490003ED0A5267536F6C7574... |

When we retrieve images, the system makes the inverse; it converts the array into an image using a class "Drawing" (In .Net), which takes the array and converts it.

In our system, images are retrieved into a grid view as in the Fig. 18. After selecting the images to be backed up, each image is converted into a hexadecimal array; this array is saved in the XML File.

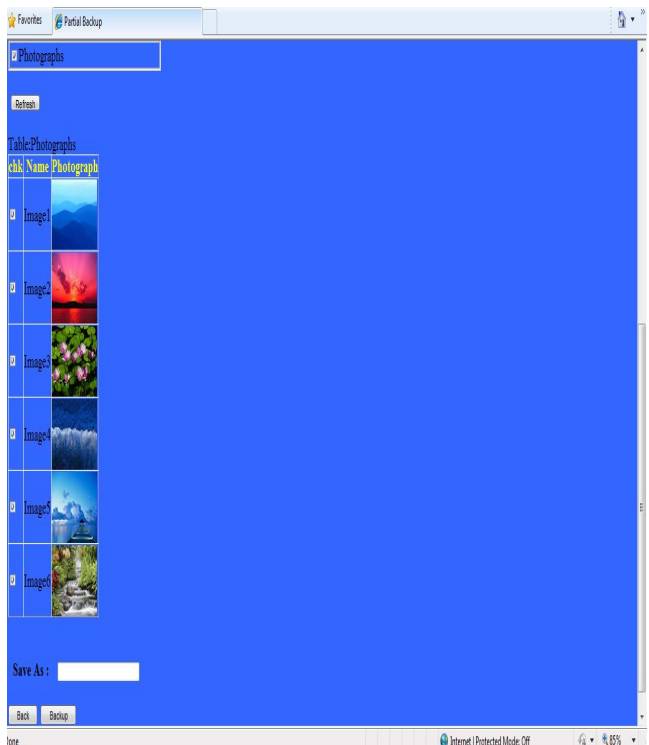

Fig. 18: Images in our Solution

The problem we faced is in the restoring operation. We could not insert the image as hexadecimal array into the database using the "insert SQL command" (this can be done only from the query analyzer). The solution of this problem (which we don't implemented) is to get the hexadecimal array and convert it into image saved in a specific directory, and then insert this image into the database. This is the available solution but it's not efficient!

## VI. CONCLUSION

In this article, we have designed and implement a new technique for Backup and Restore for specific article(s) using .NET and XML technology. Following the main points that we have achieved:

- Partially backup and restore data articles from different DBMS.
- Provide a unique friendly user interface which allows a full backup and restore of different DBMS without any need to run these systems.
- The Administrator can now backup and restore Databases whenever, whatever and wherever.

Some improvements can be added in the future in order to make this system more powerful for backup and restore:

- Add a task schedule: to automatically backup our databases after a given period.
- Make the system support more and more DBMSs as Sybase, Delphi, db2, O2…
- Implement an efficient solution to restore images into the database (different from that presented in the previous section).

- In the backup operation, user selects items to back them up. We can apply this in the restore operation to get items from the backup file and show them to the user who will select items to be restored.
- In all DBMS, and also in our solution, the restore operation clears the database and insert into it the backup file, we can provide new restore options as following :
  - Regular restore: do as above (clear and replace).
  - Merge restore: This will merge the main database by the backup file. It doesn't clear the database, but compares it by the backup file, and replaces only the data articles that hold the same name. Suppose we have a very big database of size Terra Bytes, and we are working in only one table of this database, and we need to save this table beside to restore it again if some data is lost. We can partially backup this table into XML file, if data is corrupted, we use merge restore, which does not clear our database, but replaces the table by that is in the XML file!

- Security: SQL Server 2005 save the backup files as encrypted files with extension ".BAk", our solution saves them as XML file, so we should use some encryption functions to save the backup XML file as encrypted file. In the restore operation we decrypt the encrypted file into XML file.
- Further additions: add new data articles to our partially backup operation (as users, roles, default…).
- To avoid replication of the name of the output backup file, we can make the file name of the following format: "DBMS_DBName_Date_Time"; for example: SQL_Users_13-08-2009_14.30. paradigm